\documentclass[a4paper,fleqn]{cas-sc} 

\usepackage[numbers,sort&compress]{natbib}

\usepackage[version=4]{mhchem} 
\usepackage{siunitx} 
\usepackage{gensymb} 
\usepackage{amsmath} 
\usepackage[section]{placeins} 
\usepackage[capitalise,nameinlink]{cleveref} 
\crefformat{figure}{\textbf{#2Figure~#1#3}}
\Crefformat{figure}{\textbf{#2Figure~#1#3}}
\crefformat{table}{\textbf{#2Table~#1#3}}
\Crefformat{table}{\textbf{#2Table~#1#3}}

\begin{document}
\let\WriteBookmarks\relax
\def\floatpagepagefraction{1}
\def\textpagefraction{.001}

\shorttitle{Hyperbolic metamaterial Janus particles}    

\shortauthors{S. Kattekola et al.}  

\title [mode = title]{Silver/alumina-capped Janus particles: a dispersible hyperbolic metamaterial}  



%

\author[1]{Samhita Kattekola}
\credit{Conceptualization, Data curation, Formal analysis, Investigation, Methodology, Validation, Visualization, Writing - original draft, Writing - review \& editing}
\affiliation[1]{organization={Department of Chemical Engineering, The City College of New York, CUNY},
            addressline={}, 
            city={New York},
            postcode={10031}, 
            state={NY},
            country={USA}}
            
\author[2]{Raymond R. Dagastine}
\credit{Methodology, Resources, Formal analysis, Supervision, Writing - review \& editing}
\affiliation[2]{organization={Department of Chemical Engineering, University of Melbourne},
            addressline={}, 
            city={Parkville},
            postcode={3010}, 
            state={VIC},
            country={Australia}}

\author[1]{Alexander Couzis}
\credit{Resources, Formal analysis, Supervision, Writing - review \& editing}

\author[1]{Ilona Kretzschmar}
\cormark[1]
\ead{kretzschmar@ccny.cuny.edu}
\credit{Conceptualization, Methodology, Investigation, Formal analysis, Funding acquisition, Project administration, Resources, Supervision, Writing - review \& editing}
\cortext[1]{Corresponding author}


\nonumnote{Abbreviations: AgJP, silver-capped Janus particle; HMM, hyperbolic metamaterial film; HMMJP, hyperbolic metamaterial-capped Janus particle; RIM, resonance imaging microscopy; 4P-HMM, four-period silver/alumina hyperbolic metamaterial.}

\begin{abstract}
Hyperbolic metamaterials (HMMs) made of alternating layers of metal and dielectric have been heralded as promising materials for super-resolution imaging and spontaneous emission enhancement sought after in biological applications.
However, the practical utilization of HMMs has been limited, primarily by the solid and stiff nature of the underlying substrate, usually a silicon wafer or glass substrate.
An ability to transfer HMMs to colloidal substrates would introduce dispersibility to HMMs and potentially enable their use in medical applications while extending the curved HMM geometry to discrete particles.
Here, spherical particles with alternating metal and dielectric layers are used to fabricate hyperbolic metamaterial-capped Janus particles (HMMJPs).
Alternating layers of silver (metal) and alumina (dielectric) films are evaporated onto polystyrene particle monolayers with and without fluorescent properties to study the impact of particle curvature on the layer structure in the cap.
The thickness and number of layers are chosen such that the final cap film resembles the structure of a 4P-HMM film, four periods of alternating silver/alumina, previously characterized as an HMM at wavelengths above \qty{403}{\nm}.
The fluorescent character of the polystyrene particles is used to illuminate the pathway of light in the HMMJP clusters as a function of cluster size and geometry. 
Subsequently, resonance imaging microscopy (RIM) with the HaloElement is used to independently verify the HMM nature of a single HMMJP. 
\end{abstract}


\begin{highlights}
\item Novel Janus particles with a hyperbolic metamaterial cap.
\item Hyperbolic metamaterial-capped Janus particles introduce dispersibility to metamaterials.
\item Fluorescence and resonance imaging microscopy identify the hyperbolic nature of the Janus particles with hyperbolic metamaterial caps.
\end{highlights}


\maketitle

\section{Introduction}
Hyperbolic metamaterials (HMMs) are uniaxial anisotropic materials, whose effective electric permittivity tensor has opposite signs in the parallel and perpendicular direction, allowing for hyperbolic dispersion of light.~\cite{poddubnyHyperbolicMetamaterials2013,shekharHyperbolicMetamaterialsFundamentals2014,ramirez-aragonTunabilityPlasmonicResonances2024}
As such, HMMs have the potential to be disruptive in super-resolution imaging~\cite{liuFarFieldOpticalHyperlens2007} and emission enhancement applications.~\cite{sreekanthLargeSpontaneousEmission2014}
The most common realization of HMMs is a multilayered structure made of metal and dielectric layers that is anisotropic along the z-axis on a solid substrate. 
However, the planar, substrate-bound HMM system creates a practical mismatch with many of the proposed applications, often in biological fields, which require dispersibility for operation in a liquid environment.~\cite{kimOpticalMetasurfacesBiomedical2025, sreekanthExtremeSensitivityBiosensing2016, vasilantonakisRefractiveIndexSensing2015, shafferUtilizingPowerCerenkov2017,abad-arredondoFluorescenceTriggeredRadioactive2020, zafarEvaluatingHyperbolicDispersion2023}

Colloidal metamaterials provide a route to address the limitation of a substrate-bound system by translating the metamaterial properties to a dispersible particle-based system.~\cite{huhExploitingColloidalMetamaterials2020}.
Extending HMMs to a colloidal architecture could therefore provide the dispersibility required for operation in liquid environments while retaining the anisotropic optical properties associated with the metal-dielectric multilayer structure.
Most colloidal metamaterials to date have been realized via template-assisted colloidal lithography to create particles which produce a metamaterial response when aligned to certain templates.
Several shapes, structures, materials (both organic and inorganic), and assemblies have been theorized, fabricated, and reviewed for utility in a variety of applications.~\cite{leeMetalNanoclusterMetamaterial2009, robertsMultipleTunableHyperbolic2020,rybinMetaphotonicsSubwavelengthDielectric2024, tammaVisibleFrequencyMagnetic2010, kadic3DMetamaterials2019, LiRefractoryPlasmonics2014, ebrahimiMagneticModeCoupling2023}
In particular, the translation of the hyperbolic metamaterial response to a colloidal architecture has relied on deposition of layered structures into nano-cavities, resulting in an antenna- or grating-like HMM array assembly.~\cite{zhaoHyperbolicNanoparticlesSubstrate2021, zhaoHyperbolicMetamaterialNanoparticles2023, dhamaUnveilingLongLivedHotElectron2023, maccaferriHyperbolicMetaAntennasEnable2019, maccaferriEnhancedNonlinearEmission2021, kuttruffMagnetoOpticalActivityNonmagnetic2021, kochanowskaDualpeakSensingHyperbolic2026, indukuriWS2MonolayersCoupled2020, maccaferriDesignerBlochPlasmon2020}
Other shapes such as square-cuboid particles~\cite{tangVersatileMultilayerMetamaterial2020} and lens-like arrays~\cite{byunDemonstrationNanoimprintedHyperlens2017} have also been demonstrated.
Additionally, core-shell cylinders, ~\cite{kimSubwavelengthCoreShell2017} which display hyperbolic dispersion with a modified geometry~\cite{moradiTheoryElectrostaticWaves2023} have been fabricated. 
However, the demonstrated metamaterial behavior in most of these materials has either been upon assembly or in a templated array, and therefore fluid-like dispersibility is not achieved.

There are two published works thus far that are potential candidates for a dispersible HMM.
The works of Wang et al.~\cite{wangMetaparticlesDressingNanoObjects2018} and Lin et al.~\cite{linNanoscaleCoreShell2019} introduce chemically synthesized core-shell metal-dielectric particles that can encode hyperbolic dispersion.
Both of them design HMMs with a radial effective component,~\cite{moradiTheoryElectrostaticWaves2023} which has been theorized to support resonant modes identical to whispering-gallery modes,~\cite{wuSecondHarmonicGeneration2017,wanLowThresholdSpaser2017,wuElectrodynamicalLightTrapping2014,guUltralargeRabiSplitting2021,guMultipleSharpFano2021,baqirQuasistaticAnalysisHyperbolic2021,moradiRevisitQuasistaticAnalysis2023} specifically shown to be useful in bioimaging applications.~\cite{abad-arredondoFluorescenceTriggeredRadioactive2020}
While chemical synthesis of core-shell particles by themselves has been well established,~\cite{ghoshchaudhuriCoreShellNanoparticles2012,jankiewiczSilicaMetalCore2012,hanskeSilicaCoatedPlasmonicMetal2018} several layers or shells are required to develop the HMM characteristic in a core-shell particle. 
The synthesis of such particles, therefore, is non-trivial and the necessary chemical control required with increasing number of layers creates difficulty in customizing the HMMs for different operating wavelengths or applications.

\begin{figure}
 \centering\includegraphics[width=\linewidth]{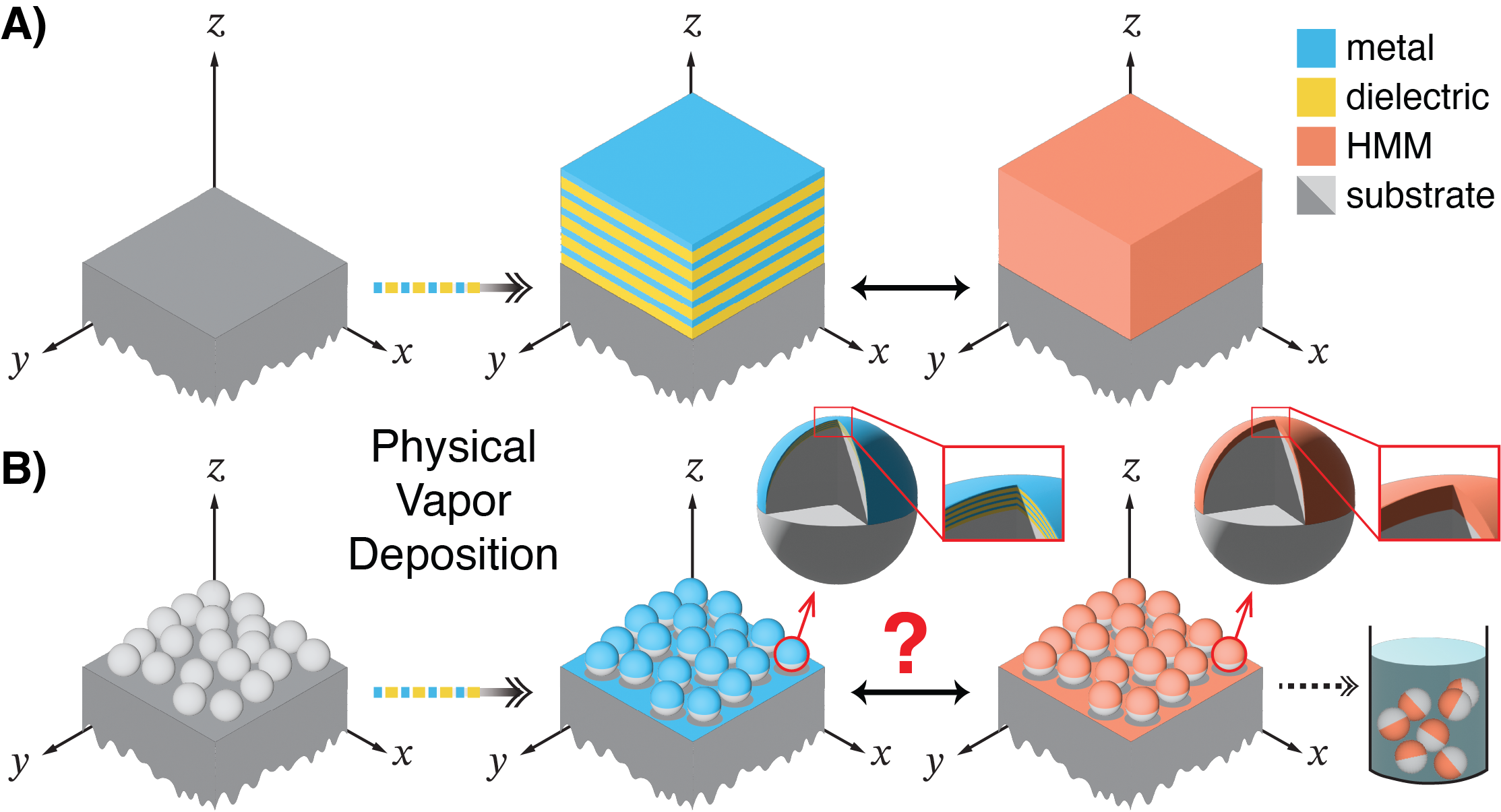} 
  \caption{Schematic showing the fabrication of HMMs. A) Traditional fabrication of a multilayered metal-dielectric film using  physical vapor deposition that behaves as a uniaxial HMM film. B) Same fabrication process with a microparticle-coated substrate yielding Janus particles with the same layered structure highlighting the question investigated here: Does the layered cap structure retain its HMM character in the Janus particle configuration?}
\label{fig:intro}
\end{figure}

\cref{fig:intro} first introduces the traditional method of making alternating layered metal (blue) and dielectric (yellow) structures that exhibit HMM character (illustrated by orange color) using physical vapor deposition onto a solid substrate (gray) in \cref{fig:intro}\textbf{A}.
\cref{fig:intro}\textbf{B} displays an alternative method by proposing the same process on a substrate with a microparticle layer, resulting in Janus particles with HMM caps, so-called HMMJPs.
HMMJPs are a simpler, intermediate solution to chemically synthesized, dispersible HMMs with a radial effective component.
Line-of-sight physical vapor deposition is the tool of choice for the fabrication of Janus particles with metal and metal oxide caps,~\cite{waltherJanusParticlesSynthesis2013,huFabricationPropertiesApplications2012} and is also compatible with the sequential fabrication required for multilayer materials. 
For example, magnetic multilayer caps have been extensively fabricated and characterized using this approach,~\cite{albrechtMagneticMultilayersNanospheres2005,kimlingneemoserMagnetoresistiveEffectsCo2010} with Ulbrich et al. demonstrating Co/Pd caps containing up to 80 bilayers.~\cite{ulbrichEffectMagneticCoupling2010}
Lee et al. used alternating \ce{SiO2}/\ce{TiO2} layers to produce magnetoresponsive photonic Janus particles.~\cite{leeMagnetoresponsivePhotonicMicrospheres2017}

Our work inquires whether a film of alternating silver and alumina layers with four periods grown onto a colloidal microparticle retains its metamaterial character yielding a dispersible metamaterial (\cref{fig:intro}\textbf{B}). 
For this purpose, colloidal monolayer clusters of polystyrene microparticles with and without fluorescence properties are prepared on glass substrates.
An HMM layer is coated onto particle monolayer clusters and subsequently, interrogated with various tools such as focused ion beam-scanning electron, bright-field, epifluorescence, and resonance imaging microscopy (RIM). 
Here, we show that the layered HMM structure successfully transfers to the particle, resulting in a cap with an HMM structure, and that the cap exhibits metamaterial character. 

\section{Materials and methods}

\textbf{Sample Preparation and Material Fabrication.} 
Glass slide (Fisher Scientific), single-side polished silicon wafer (University Wafer, Inc.), and round glass waveguide (G-S Supplies, Inc.) substrates are cleaned by submerging them in a mixture of Nochromix\textregistered\, and concentrated sulfuric acid (Fisher Scientific) for two hours, followed by copious rinsing with deionized water (Millipore, \qty{18.2}{\mega\ohm\centi\metre}) and drying in an oven at \qty{70}{\celsius}. 
\qty{2.4}{\um} sulfated polystyrene particles (\qty{8}{\percent} w/v) and \qty{2.0}{\um} red fluorescent microspheres (\qty{1}{\percent} solids, Firefli Fluorescent Red dye: \qty{542}{\nm}/\qty{612}{\nm}) purchased from Fisher Scientific are used in the experiments.
\qtyrange{60}{100}{\uL} of solution comprising water and polystyrene particles with a concentration of $\sim$\qty{3}{\mg\per\mL} is dispersed via controlled convection~\cite{prevoControlledRapidDeposition2004} onto the substrates to form submonolayer particle assemblies, referred to as particle monolayer substrates, comprising isolated particles and clusters.
A cluster is defined as an aggregation of at least two particles.
E-beam evaporation (AJA International Orion 8E Evaporator) is used to deposit four periods of alternating layers (nominal thickness) of alumina (\qty{15}{\nm}), germanium (\qty{2}{\nm}), and silver (\qty{10}{\nm}) terminated by an alumina protection layer (\qty{6}{\nm}) creating particle monolayer substrates with an actual \qty{128}{\nm} 4P-HMM coating in accordance with our earlier work.~\cite{Kattekola}
The germanium serves as a seed layer to promote smooth silver layer growth ~\cite{logeeswaranvjUltrasmoothSilverThin2009} and cannot be distinguished from the silver layer.~\cite{Kattekola}
During evaporation, both the particles and the substrate are coated leading to 4P-HMM Janus particles surrounded by a 4P-HMM film, hereafter referred to as HMM-HMMJPs.
After deposition, HMM-HMMJPs on glass slides are used as is for microscopy characterization, while those on silicon wafers are used for structural characterization of the 4P-HMM cap. 
The same process is followed to create particle monolayer substrates with caps of germanium-seeded silver terminated with an alumina protection layer (nominal thicknesses: \qtylist[list-final-separator={, and }]{2;40;6}{\nm}, respectively) as the control sample, which has the same total amount of silver as the 4P-HMM, hereafter referred to as Ag-AgJPs.

\textbf{Focused Ion Beam - Scanning Electron Microscopy.} 
An FEI Helios NanoLab 660 focused ion beam - scanning electron microscope (FIB-SEM) system is used for both precise ion-milling and high-resolution imaging to characterize the structural integrity of the cap following the approach published by Rashidi et al.~\cite{Rashidi} using an isolated particle on a silicon wafer. 
Next, in a rectangular area surrounding the particle of interest, platinum is sequentially deposited in \qty{100}{\nm} increments to yield a total layer thickness of \qty{500}{\nm} using the electron beam-induced gas injection system.
Using a \ce{Ga+} ion beam, approximately \qty{50}{\percent} of the particle is milled off. 
SEM images of the particle cross section after platinum deposition and milling as well as close-up images of the various cap regions are taken at \qty{52}{\degree} tilt with respect to the substrate. 

\textbf{Bright-field and Fluorescence Microscopy.} 
An Olympus BX-51 microscope with an Olympus SLMPlan N100x super long working distance objective and an AmScope 9.0 MP CMOS sensor camera is used for bright-field imaging in reflection and transmission mode. 
Transmission fluorescence microscopy is performed using the Olympus BX-51 microscope in transmission mode with the condenser removed and eyepieces disabled. 
A generic laser pointer of wavelength \qty{532}{\nm} (brand: XIMBI, purchased from Amazon) mounted using a clamp is pointed at the sample from below the stage. 
A \qty{590}{\nm}, \qty{12.7}{\mm} diameter, colored glass long-pass filter (Edmund Optics) is added to the camera to exclude any stray laser light from being recorded.
Standard epifluorescence microscopy in inverted configuration is performed with the same objective and camera using an AmScope IN300T-FL microscope equipped with a mercury lamp with an excitation filter of \qtyrange{510}{560}{\nm} and an emission filter of \qty{590}{\nm} to image the red fluorescent particles.
Images are taken at constant camera exposure time of \qty{25}{\ms} and a gain of 1 under identical microscope light settings for the respective microscopes.

\textbf{Reflectance and Transmittance Measurements.} 
A J.A. Woollam Co., Inc. V-VASE Ellipsometer equipped with WVASE software~\cite{WVASE} is used to collect transmittance data on flat 4P-HMM and Ag films at \qty{0}{\degree} incident angle between \qty{300}{\nm} and \qty{800}{\nm} with \qty{10}{\nm} step size.
A baseline scan, with the same incident angle and data measurement range, is performed at the beginning of every run without any sample in the pathway. 
Filmetrics F20 is used to perform reflectance measurements between \qty{300.37}{\nm} and \qty{800.4}{\nm} at an $\approx$\qty{0.8}{\nm} step size with the light at \qty{0}{\degree} incident angle, i.e., normal to the surface of the film. 
A baseline measurement is performed by first placing the respective sample to capture saturation limits and then using a BK7 glass standard.

\textbf{Resonance Imaging with HaloElement.}
The HaloElement microscope attachment by Tiny Bright Things is used for RIM.~\cite{Avinash, Bolton2019}
Images are taken with the HaloElement on an Accuscope EXC-500 microscope equipped with a FLIR Grasshopper\textregistered 3 High Performance USB 3.0 Monochrome Camera using an Olympus SLMPlan N50x super long working distance objective with 0.35 numerical aperture. 
The measuring principle of the HaloElement is detailed in Section S1 and Figure S1 (Supplementary Material). 
In brief, the HaloElement operation is based on total internal reflection of LED light inside a circular glass waveguide that generates an evanescent wave at the waveguide/air interface, which scatters at the edge of a particle located on the waveguide forming an intense primary and a less intense secondary hotspot. 
Since the hotspots are the result of the evanescent wave resonating within the particle, the scattering hotspots are sensitive to the optical environment of the particle surface including coatings on the particles.
The two hotspots generated from the scattering are captured by the camera generating an image.
The HaloElement is equipped with 15 red LED lights operating at \qty{632}{\nm} and azimuthally equidistantly located along the perimeter of the waveguide such that upon illumination, scattering takes place in 15 spots along the particle perimeter. 
The images corresponding to the 15 LEDs are overlaid to obtain a single image representing the edge and scattering profile with respect to the surface normal of the particle on the waveguide.
Samples fabricated on waveguides are imaged at constant exposure time and gain to allow for direct comparison of material properties between 4P-HMM and Ag samples.
Exposure time, which refers to the LED illumination time, is maintained at \qty{4.8}{\ms} and gain is maintained at \qty{35}{\dB}.

\begin{figure}[]
 \centering\includegraphics[width=\linewidth]{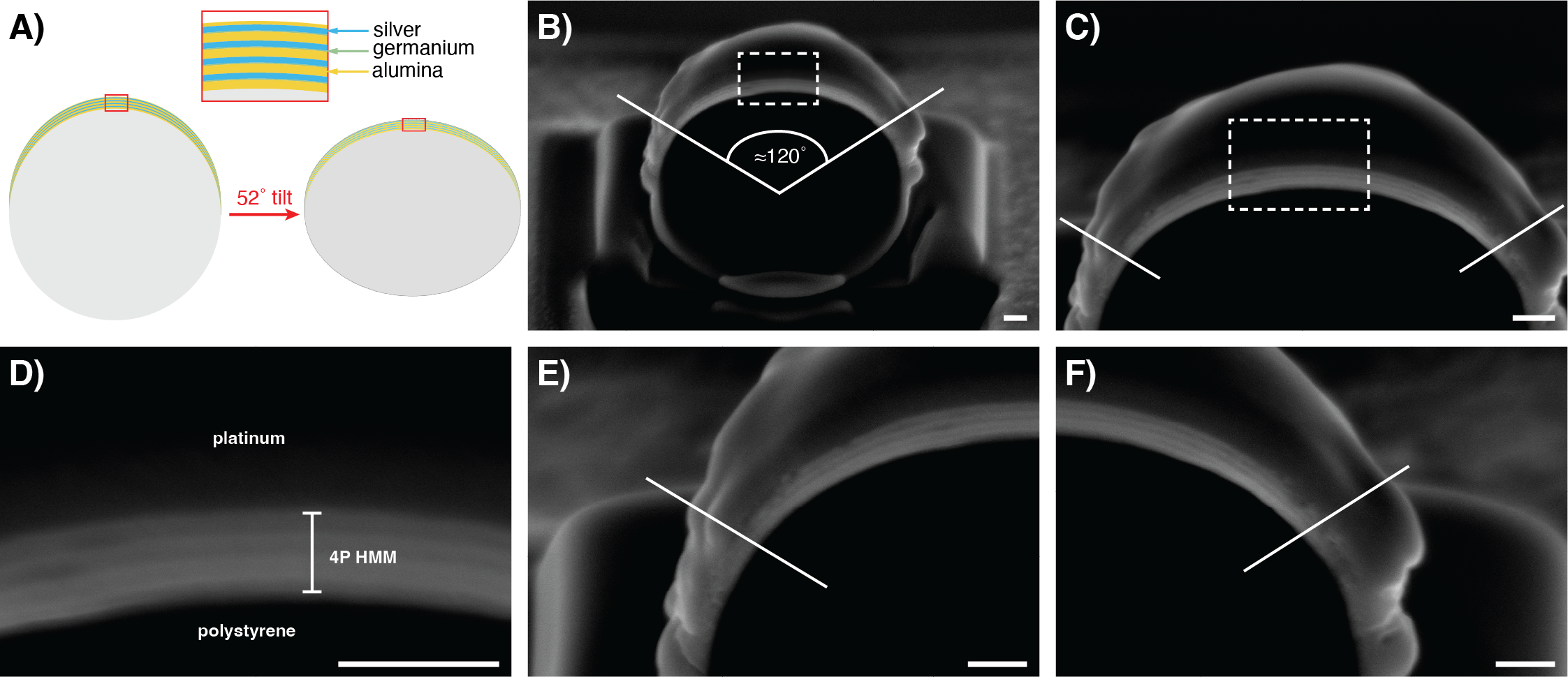} 
  \caption{Cross-sectional characterization of 4P-HMM cap of HMM-HMMJP with FIB-SEM. 
  \textbf{A)} Schematic of HMM layered structure and illustration of \qty{52}{\degree} tilt at which SEM images are taken.
  \textbf{B)} HMMJP coated with \qty{500}{\nm} \ce{Pt} protection layer and milled halfway using \ce{Ga+} FIB. 
  White lines enclose a sector of the cap ($\approx$\qty{120}{\degree}) in which all layers of the HMM are discernible. 
  High-resolution images of \textbf{C)} entire HMM cap showing the typical crescent-moon shape, \textbf{D)} zoomed-in section of cap highlighted by white dashed box in \textbf{B)} and \textbf{C)} showing nine discernible layers. The reported 4P-HMM thickness (\qty{128 +- 1}{\nm})~\cite{Kattekola} is marked on top of the actual HMM cap sandwiched between the polystyrene particle and \ce{Pt} protection layer. 
  \textbf{E)} Left and \textbf{F)} right side of cap edge. 
  Scale bars are \qty{200}{\nm}.
  }
\label{fig:FIB-SEM}
\end{figure}

\section{Results and discussion}
Janus particles are fabricated into monolayer clusters of various sizes with 4P-HMM and Ag caps.
The cap structure of individual Janus particles is analyzed using FIB-SEM to shed light on the impact of the curved particle surface on the HMM layer structure.
A combination of bright-field microscopy, epifluorescence imaging, and RIM using the HaloElement is employed to show that the HMM cap retains its metamaterial character in a single Janus particle.

The line-of-sight nature of physical vapor deposition enables deposition of layered films onto curved surfaces through use of particle substrates such as micrometer-sized polystyrene particles.
Mass conservation requires the layer structure of the film to stay intact in the line-of-sight direction; however, the layer thicknesses in the resulting cap will vary in the radial direction due to the surface curvature of the particle, demonstrated previously by Rashidi et al.~\cite{Rashidi} for a metal cap.
Here, the cap shape and layer structure of an HMM film on a \qty{2.4}{\um} sulfated polystyrene particle are characterized using cross-sectional analysis with FIB-SEM summarized in \cref{fig:FIB-SEM}.  
\cref{fig:FIB-SEM}\textbf{A} shows the schematic of the expected 4P-HMM structure on the particle to scale.
In the FIB-SEM,  the \ce{Ga+} ion beam used for milling is positioned normal to the substrate, whereas the SEM detector is located at a \qty{52}{\degree} angle, which leads to an apparent compression of the cap and particle in the surface normal direction, as illustrated in \cref{fig:FIB-SEM}\textbf{A}.
\cref{fig:FIB-SEM}\textbf{B} shows an SEM cross section of a \qty{2.4}{\um} HMMJP that has been milled halfway after application of a protective \qty{500}{\nm} platinum coating.
Additional details of the milling process are provided in the Materials and methods section and the Supplementary Material Section S2.
The white lines and dashed box in \cref{fig:FIB-SEM}\textbf{B} mark two significant features of the cap cross section, i.e., the nearly \qty{120}{\degree} sector of the particle cap cross section that shows the expected layer structure (carried through to \cref{fig:FIB-SEM}\textbf{C}, \textbf{E}, and \textbf{F}) and the area of the cap depicted in the high-resolution image provided in \cref{fig:FIB-SEM}\textbf{D}, respectively. 
\cref{fig:FIB-SEM}\textbf{C} shows a zoomed-in SEM image of the entire HMM cap, confirming the overall expected crescent-moon cap shape where the cap is the thickest at the pole and becomes thinner as it approaches the particle equator.
The dashed-boxed area, which encompasses the pole of the particle and includes the center of the cap, shown in \cref{fig:FIB-SEM}\textbf{D}, reveals nine discernible layers of alternating contrast. 
The previously reported \qty{128 +- 1}{\nm} thickness of a planar 4P-HMM film using atomic force microscopy (AFM) ~\cite{Kattekola} after correction for the \qty{52}{\degree} tilt~\cite{Rashidi} (see Section S2 in Supplementary Material) is highlighted in \cref{fig:FIB-SEM}\textbf{D} and closely matches the apparent thickness at the center of the cap.  
\cref{fig:FIB-SEM}\textbf{E} and \textbf{F} show the zoomed-in left and right quarters of the particle, respectively. 
The layered structure becomes less discernible beyond the white lines toward the edges and is likely a result of limited SEM resolution.

\begin{figure}
    \centering
    \includegraphics[width=\linewidth]{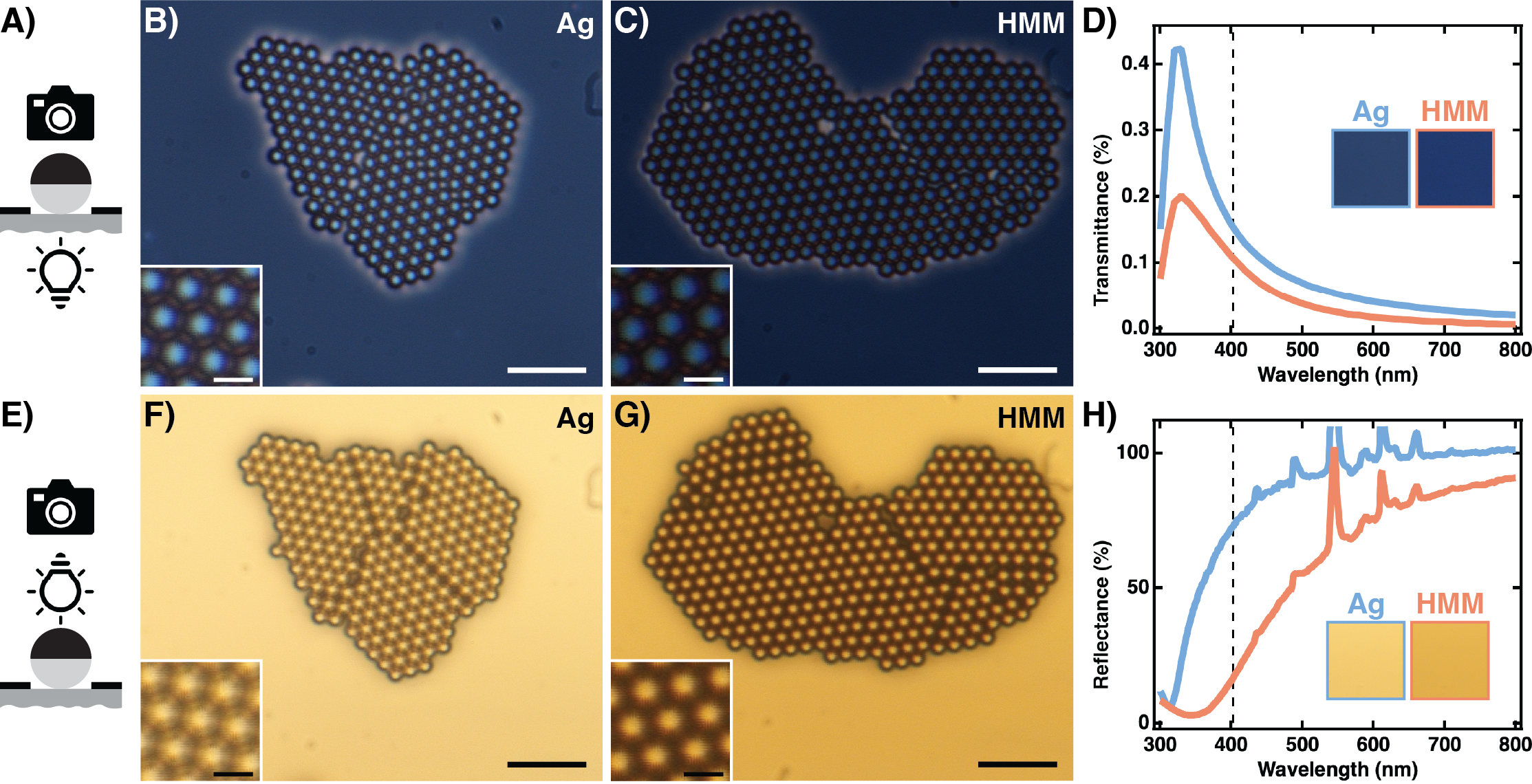}
    \caption{Bright-field characterization of Ag-AgJP and HMM-HMMJP prepared from \qty{2}{\um} fluorescent polystyrene particles in transmission \textbf{A}-\textbf{D} and reflection mode \textbf{E}-\textbf{H}.
    \textbf{A)} and \textbf{E)} show the schematic of imaging modality. \textbf{B)} and \textbf{C)} show images of Ag-AgJP and HMM-HMMJP, respectively, in transmission mode.
    \textbf{D)} Transmittance (in \%) as a function of wavelength for Ag (blue) and HMM films (orange) without particles. 
    Insets show representative images of films. The dashed lines indicate $\lambda_\mathrm{HMM} \approx \qty{403}{\nm}$ for this combination.~\cite{Kattekola} 
    \textbf{F)} and \textbf{G)} show images of Ag-AgJP and HMM-HMMJP, respectively, in reflection mode. 
    Insets show zoomed-in portions of the clusters.
    \textbf{H)} Reflectance (in \%) as a function of wavelength for Ag (blue) and HMM films (orange) without particles. 
    Insets show representative images of films. 
    Scale bars are \qty{10}{\um}. Features in the reflectance measurement are caused by ambient light.
}\label{fig:bright-field}
\end{figure}

Bright-field microscopy in reflection and transmission mode is used to characterize the optical differences between HMM-HMMJP and Ag-AgJP, in known orientation, compared to their thin-film analogues, HMM and Ag (\cref{fig:bright-field}). 
All images are acquired under identical illumination and camera settings and allow differences in the recorded color and brightness to be attributed primarily to differences in the optical properties of the HMM and Ag films.
The top row of \cref{fig:bright-field} represents transmission data whereas the bottom row shows reflection data for the same Ag-AgJP and HMM-HMMJP clusters. 
Careful inspection of the transmission images, \cref{fig:bright-field}\textbf{B} and \textbf{C}, and reflection images, \cref{fig:bright-field}\textbf{F} and \textbf{G}, reveals that the HMM film appears darker than the Ag film in both cases. 
A similar difference is observed for the Ag-AgJP and HMM-HMMJP clusters in both images (see insets).
In addition to the reduction in transmission and reflection for the HMM sample, there is an apparent glow at the edges of the clusters in transmission mode and a darker border in reflection mode.
Although the glow and darker border appear in both systems, they are more pronounced in the Ag system. 
A feature unique to the Ag system is the appearance of radially extending spokes from the centers of the AgJPs in the reflection mode image (see inset of \cref{fig:bright-field}\textbf{F}).
In contrast, the HMMJPs have darker boundaries at the rim (see inset in \cref{fig:bright-field}\textbf{G}).

The transmittance and reflectance of planar HMM and Ag films measured with light incident and collected normally to the surface of the sample (\qty{0}{\degree} angle of incidence) in \cref{fig:bright-field}\textbf{D} and \textbf{H}, respectively, quantitatively confirm the observed reduction in transmission and reflection for the HMM film. 
The insets in \cref{fig:bright-field}\textbf{D} and \textbf{H} are representative images of the planar films, while the dashed line indicates $\lambda_\mathrm{HMM} \approx \qty{403}{\nm}$, the wavelength beyond which the material turns into HMM for the silver/alumina combination used here.~\cite{Kattekola}
A steeper increase in reflectance is observed past $\lambda_\mathrm{HMM}$ for the HMM film.
Additional evidence of stronger reflectance of the Ag film is provided by the enhancement of ambient light peaks in the Ag film reflectance data compared to HMM (\cref{fig:bright-field}\textbf{H}). 
The reflection and transmission trends seen in \cref{fig:bright-field}\textbf{D} and \textbf{H} are in agreement with measurements reported by Tumkur et al. on \ce{Ag/MgF2} metamaterial and \ce{Ag} films.~\cite{tumkurControlReflectanceTransmittance2012}

\begin{figure}
    \centering
    \includegraphics[width=\linewidth]{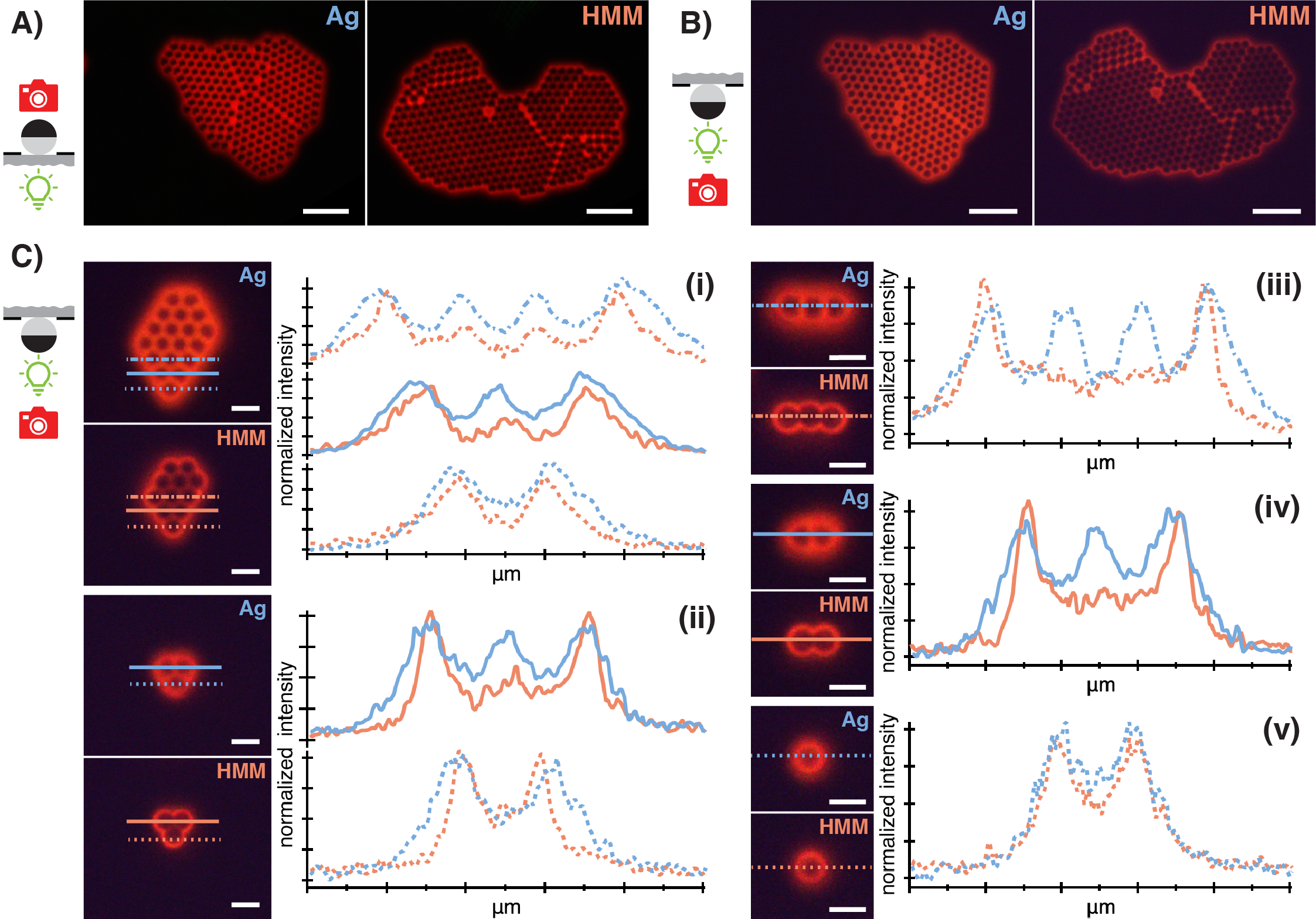}
    \caption{Fluorescence characterization of Ag-AgJP and HMM-HMMJP prepared from \qty{2}{\um} fluorescent polystyrene particles. 
    \textbf{A)} Big clusters of AgJP (208 particles) and HMMJP (349 particles) are used to detect fluorescence through the cap using a transmission configuration. 
    \textbf{B)} Images of the clusters shown in \textbf{A)} using an epifluorescence configuration exhibiting identical features. 
    \textbf{C)} Epifluorescence imaging of \textbf{i)} small clusters (19 AgJP and 11 HMMJP), \textbf{ii)} triplets, \textbf{iii)} chains, \textbf{iv)} doublets, and \textbf{v)} single particles accompanied by intensity profiles through three (dot-dashed line), two (solid line), and one particle (dotted line) for AgJP (blue) and HMMJP (orange).
    Scale bars in \textbf{A)} and \textbf{B)} are \qty{10}{\um} and \qty{3}{\um} in \textbf{C)}.
    }
    \label{fig:epifluorescence}
\end{figure}

Having established that the transmission and reflection differences in HMM and Ag films transfer to the  Ag-AgJP and HMM-HMMJP clusters, the optical response of the clusters fabricated from red fluorescent \qty{2}{\um} particles is investigated further using fluorescence microscopy (\cref{fig:epifluorescence}). 
\cref{fig:epifluorescence}\textbf{A} and \textbf{B} show fluorescence emission and epifluorescence emission, respectively, through the cap for the same two Ag-AgJP and HMM-HMMJP clusters shown in \cref{fig:bright-field}.
In both HMM-HMMJP cluster images (right image in \cref{fig:epifluorescence}\textbf{A} and \textbf{B}), individual HMMJPs inside the cluster show dark rims, while cluster edges and grain boundaries are bright. 
In contrast, in the Ag-AgJP cluster images the individual particle rims, cluster edges, and grain boundaries have comparable brightness.
Further, the Ag-AgJP cluster appears to have a more diffuse edge.
Both observations indicate distinct differences between the light interaction in Ag-AgJP and HMM-HMMJP clusters. 

To investigate this behavior further, epifluorescence emissions of AgJP and HMMJP clusters of smaller sizes down to a single particle are collected in  \cref{fig:epifluorescence}\textbf{C} together with intensity profiles for several lines superimposed on the epifluorescence images using Fiji.~\cite {schindelinFijiOpensourcePlatform2012}
Since the amount of dye in an individual particle needed for quantification of particle emission is unknown, all images are self-normalized to evaluate only the relative differences between light emission in each image.
The trend of bright rim and dark interior observed for the big HMM-HMMJP cluster in \cref{fig:epifluorescence}\textbf{A} and \textbf{B} is retained for the small HMM-HMMJP cluster in \cref{fig:epifluorescence}\textbf{C}(i) through to the HMM-HMMJP doublet in \cref{fig:epifluorescence}\textbf{C}(iv).
Interestingly, the line scan analysis in \cref{fig:epifluorescence}\textbf{C} indicates that cluster geometry matters regarding the fluorescence uniformity inside the cluster. 
Comparison of the line scans for Ag-AgJP (blue) and HMM-HMMJP (orange) clusters in \cref{fig:epifluorescence}\textbf{C}(i) reveals a decreased intensity for the peaks (particle rims) inside the HMM-HMMJP cluster compared to the particle rims located on the outside of the cluster, while peak intensities remain nearly constant for the  Ag-AgJP clusters.  
In addition, a reduction in the cluster size from \cref{fig:epifluorescence}\textbf{C}(i) to \cref{fig:epifluorescence}\textbf{C}(ii) leads to a sharpening of the external particle rims in the HMM-HMMJP cluster. 
Surprisingly, the intensity for the internal particle rims reduces to an even smaller amount when the HMMJPs are aligned in linear chains as seen in \cref{fig:epifluorescence}\textbf{C}(iii) and \textbf{C}(iv). 
For a single HMM-HMMJP, \cref{fig:epifluorescence}\textbf{C}(v), a slight intensity reduction inside the particle persists compared to the single Ag-AgJP, i.e., the HMM cap appears to decay the amount of light coming out of the fluorescent particle more than the Ag cap, however, differences in fluorescent dye loading prevent its quantification.~\cite{watersAccuracyPrecisionQuantitative2009}  
The overall reduction of fluorescence in HMMJPs within the chains demonstrates that the HMM caps in a chain of HMMJPs behave the way a planar HMM does when the pathway for outcoupling is restricted. 

The directionality in the reduction of fluorescence in \cref{fig:epifluorescence} suggests the impact of the pathway allowed in the HMM.
The reflection decreases as the pathway between the edges of the HMM cluster increases, implying that the HMM caps absorb the fluorescence,  potentially into high-k modes. 
 At the edges of each cluster or in the single particle, the particle architecture for HMM enables scattering off the edge of the particle, thus, the light in high-k modes is not dissipated as losses.
However, when two HMMJPs are touching, as the light pathway encounters multiple HMM/scatterer interfaces, the compounded optical loss during transport causes an overall reduction in the observed reflection between particles.  
The compounding loss of reflection as a function of particle assembly is further evident in \cref{fig:epifluorescence}\textbf{B}.
In the HMM-HMMJP cluster, the darker areas in the bottom left of the cluster have more particles closely packed with no defects, whereas visually similar packing in the top right is lighter, which is further highlighted by a twinning boundary defect in the cluster lattice.
In contrast, the Ag-AgJP cluster appears evenly illuminated regardless of particle packing.

 \begin{figure}
 \centering\includegraphics[width=0.75\linewidth]{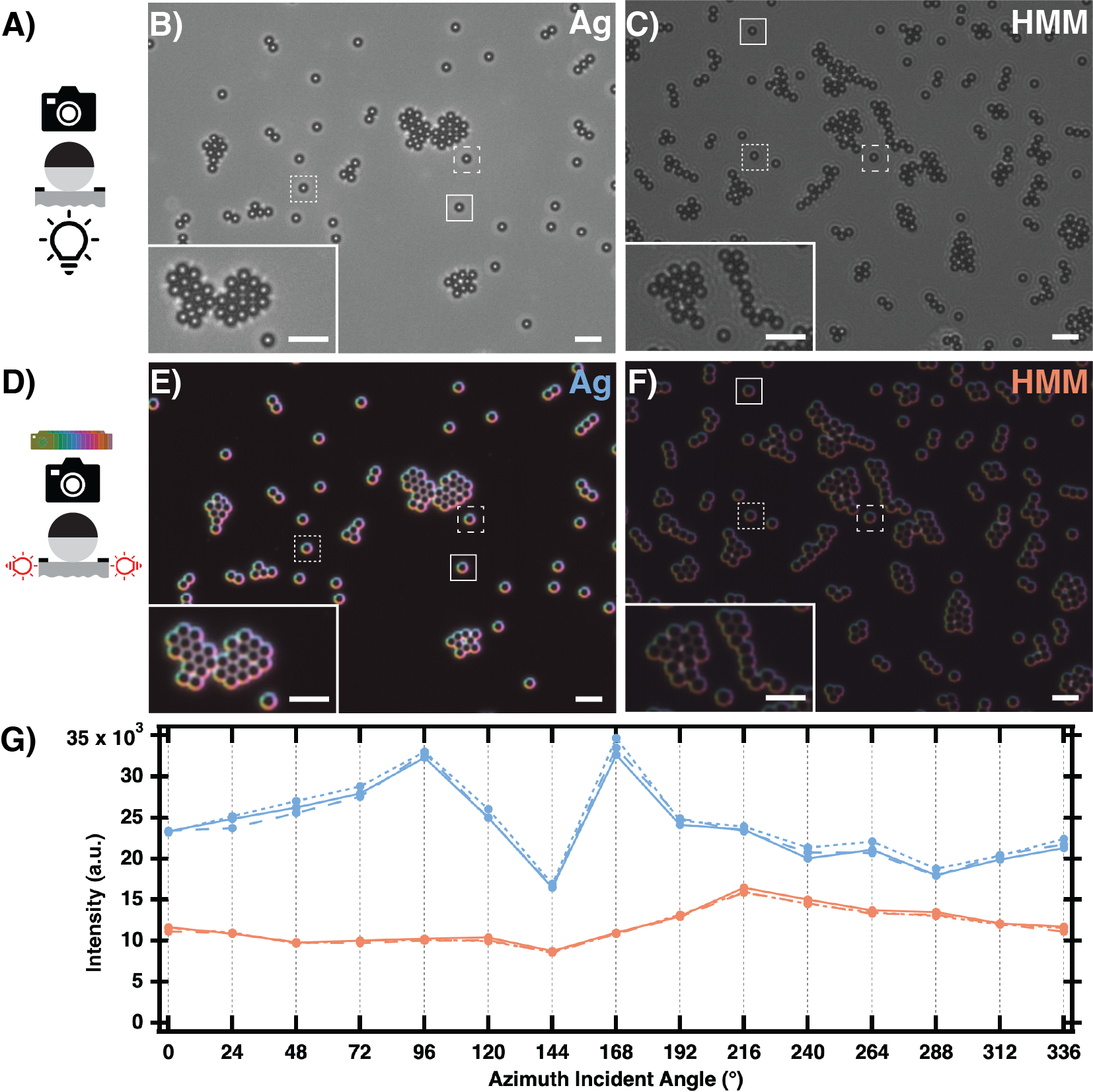} 
  \caption{Resonance imaging microscopy of Ag-AgJP and HMM-HMMJP prepared from \qty{2.4}{\um} polystyrene particles in bright-field (\textbf{A}-\textbf{C}) and resonance imaging mode (\textbf{D}-\textbf{F}). 
  \textbf{A)} and \textbf{D)} show schematics of imaging modality. 
  \textbf{B)} and \textbf{C)} show bright-field images of Ag-AgJP and HMM-HMMJP, respectively, with insets of enlarged representative clusters.
  \textbf{E)} and \textbf{F)} show composites of 15 images taken using the HaloElement for Ag-AgJP and HMM-HMMJP, respectively, with insets of the same representative clusters as in \textbf{B)} and \textbf{C)}. 
  White boxes around individual particles in \textbf{B,C)} and \textbf{E,F)} indicate particles used for intensity plots in \textbf{G)}, which shows intensity of the primary hotspot for Ag (blue) and HMM (orange) in environments with isolation (solid line), other single particles (dotted), and small clusters (dashed)  as a function of the azimuth incident angle.
  Scale bars are \qty{10}{\um}.
 }\label{fig:HaloMain}
  \end{figure}

Resonance imaging microscopy (RIM) using the HaloElement is performed to gain insight into the propagation of light in a single HMM-HMMJP (\cref{fig:HaloMain}) as well as how the propagation of evanescent wave scattering compares to the interaction of plane wave light in bright-field and fluorescence in  HMM-HMMJP clusters.
The top and middle rows of \cref{fig:HaloMain} show the bright-field and RIM images, respectively, taken with the HaloElement for Ag-AgJP and HMM-HMMJP samples prepared from non-fluorescent \qty{2.4}{\um} sulfated polystyrene particles on circular glass waveguides. 
The white boxes in the images indicate particles used for intensity plots shown in \cref{fig:HaloMain}\textbf{G}.
Since the RIM images are acquired using non-fluorescent particles, taken with the same exposure time and gain, and displayed at the same total scale, the intensity between the isolated Ag-AgJP and HMM-HMMJP can be quantitatively compared.
Similar to the bright-field transmission images in \cref{fig:bright-field}\textbf{B} and \textbf{C}, the Ag-AgJP sample appears brighter than the HMM-HMMJP sample in \cref{fig:HaloMain}\textbf{B} and \textbf{C}.
The glow at the edges of the Ag-AgJP clusters radiates as bright ripples around the clusters on the surrounding Ag film.
Conversely, the darkness at the edges of the HMM-HMMJP clusters radiates as darker ripples on the surface of the HMM planar film.
There are radially extending spokes visible in both Ag-AgJP and HMM-HMMJP clusters located at the edges of the clusters in between particles. 
In the composite RIM images (\cref{fig:HaloMain}\textbf{E} and \textbf{F}), the Ag-AgJP sample is also brighter than the HMM-HMMJP sample.
Similar to \cref{fig:epifluorescence}, every HMM-HMMJP cluster shows a reduced particle rim intensity within the cluster and brighter cluster edges and grain boundaries, whereas edges, grain boundaries, and particle rims in the Ag-AgJP clusters exhibit similar brightness.
The consistency between the fluorescence images in \cref{fig:epifluorescence} and the RIM observations in \cref{fig:HaloMain} provides fluorescence-independent evidence that the reduced light intensity observed for HMM-HMMJP clusters relative to Ag-AgJP clusters is intrinsic to light propagation within the HMM caps in the clusters. 
Furthermore, the overall texture of the HMM planar film appears more speckled than that of the Ag planar film, suggesting spatial nonuniformity in the transmission of light through the waveguide, as previously reported in the literature.~\cite{leeMetamaterialAssistedIllumination2021}
The texture also suggests enhanced scattering within the planar HMM film due to the waveguide's surface roughness.~\cite{amraTrappedLightScattering2021}

Further insight is provided by the analysis  of the individual channels for single particles highlighted using white boxes in \cref{fig:HaloMain}\textbf{B}-\textbf{C} and \textbf{E}-\textbf{F}. 
Three singlets with different environments are chosen from each sample. 
The particles boxed with a solid line are isolated, i.e., there are no other particles or clusters within four particle diameters. 
The particles boxed with the dotted and dashed lines are in proximity to other singlets and small clusters, respectively.
\cref{fig:HaloMain}\textbf{G} shows the intensity of the primary hotspot for each particle as a function of the azimuth incident angle, i.e., LED position.
The overall shape of the intensity vs. azimuth incident angle is influenced by the variability of the LEDs. 
The particle environment appears to slightly impact the intensity of both HMMJP and AgJP, with an average intensity difference per LED between the three environments of \qty{2.8}{\percent} and \qty{4.2}{\percent}, respectively.
The instrument's variability for polystyrene particles is approximately \qty{6}{\percent} per LED (see Supplementary Material, Figure S1\textbf{E}).
Thus, the fluctuations in intensity between different environments for both HMMJP and AgJP are within the instrument's LED variability. 
The intensity drop at the \qty{144}{\degree} LED relative to the adjacent LEDs for the HMMJP (\qty{18}{\percent}) is reduced when compared to the control measurements of uncapped polystyrene particles (\qty{39}{\percent}) shown in the Supplementary Material Figure S1\textbf{E} or the AgJP (\qty{43}{\percent}) in \cref{fig:HaloMain}\textbf{G} indicating that the HMM caps disperse the evanescent wave scattered light more uniformly.
Further, the channel-to-channel LED intensity for the HMMJP is reduced when compared to the individual AgJP in \cref{fig:HaloMain}\textbf{G}.
Averaging the blue readings ($(24 \pm 5)\times10^{3}$) and orange readings ($(12 \pm 2)\times10^{3}$), the average intensity of light captured during RIM for an HMM-HMMJP is reduced by \qty{51 +- 13}{\percent} compared to the Ag-AgJP for all three environments. 

Evanescent wave illumination of a particle differs from illumination of a particle in both bright-field and fluorescence imaging. 
An evanescent wave, traveling parallel to the glass surface, only penetrates into the air several hundred nanometers from the glass surface.
Thus, the evanescent wave only interacts with a small fraction of the particle that is \qty{2.4}{\um} in height.
The resulting particle illumination comes from the internal scattering of light as it bounces (i.e., transmission or reflection from multiple sequential reflections at optical interfaces~\cite{Prieve:93}) or resonates~\cite{LIU_95} inside the entire particle, scattering at both the polystyrene-air and Janus cap-air interfaces for either the silver or the HMM cap. 
The reduced scattering observed for the HMM-HMMJP case in \cref{fig:HaloMain}\textbf{G} can therefore be attributed to the modulation of evanescent wave light scattering by the HMM cap.
The overall reduction of intensity in RIM (\cref{fig:HaloMain}) implies that the fluorescence reduction in \cref{fig:epifluorescence} is caused by the hyperbolic nature of the HMM cap on a Janus particle and is independent of fluorescence.

\section{Conclusions}
HMMJPs, in a known, upright orientation, were characterized using FIB-SEM, bright-field, fluorescence, and resonance imaging microscopy and showed that the cap retained the structural and optical characteristics of the corresponding planar HMM film.
FIB-SEM verified that the multilayer architecture of the planar HMM was retained despite deposition onto a curved particle surface. 
Bright-field microscopy showed that HMMJPs exhibited opacity similar to that of planar HMM films with distinguishable optical properties from AgJPs.
Fluorescence microscopy indicated a selective reduction in fluorescence within HMM-HMMJP clusters and an enhanced reduction in linear chains explained by the restriction in the available pathways for light outcoupling. 
The selective reduction of intensity within HMM clusters was validated independently of fluorescence using RIM, where the scattering intensity of individual HMM-HMMJPs was reduced by more than \qty{50}{\percent} relative to that of individual AgJPs.
The reduction in intensity suggests the presence of high-k modes within the individual particles, indicating that the HMM cap retains the characteristics of the corresponding planar HMM and confirming the fabrication of HMMJPs. 

Overall, the work shows that the Janus-particle approach to dispersible HMM materials is valid, combining the utility of Janus particles with the optical behavior of HMMs and providing straightforward access to particles that can be used in future applications that require emission enhancement or super-resolution.~\cite{shafferUtilizingPowerCerenkov2017,abad-arredondoFluorescenceTriggeredRadioactive2020} 
Future work using HMMJPs will focus on isolating individual particles from the HMM-coated substrate to directly investigate the light-outcoupling mechanism promoted by the curvature of the microsphere supporting the HMM cap using time-resolved techniques.~\cite{galfskyActiveHyperbolicMetamaterials2015} 
While removal of the HMMJPs from the substrate is straightforward using sonication, the main difficulty with single particle measurements will be the arrangement of particles such that their cap orientation is known, however, the inclusion of magnetic layers may help address this problem.
Measurements on isolated particles could establish the influence of cap orientation on access to high-k modes.
Subsequently, the impact of particle diameter and size on the HMM cap properties would be interesting to investigate, requiring minimal modification to the fabrication process, while still retaining the simplicity of using colloidal lithography.~\cite{pawarMultifunctionalPatchyParticles2009}

\printcredits
\section*{Supplementary Material}
Supplementary Material provides the HaloElement setup and measuring principle and FIB-SEM process details.
\section*{Acknowledgements}
SK, AC, and IK acknowledge support from the NSF Phase II CREST Center for Interface Design and Engineered Assembly of Low Dimensional Materials (IDEALS), NSF grant number EES-2112550. 
SK is thankful to the Chemical Engineering Travel Fund of the ChE Department at The City College of New York for providing funds to visit the Dagastine Laboratory at the University of Melbourne.
This work was performed in part at the Materials Characterisation and Fabrication Platform (MCFP) at the University of Melbourne and the Victorian Node of the Australian National Fabrication Facility (ANFF).
IK acknowledges funding through PSC CUNY award 67759-00 55. 
This work was performed in part at the Nanofabrication Facility at the Advanced Science Research Center at The Graduate Center of the City University of New York. 
\bibliographystyle{elsarticle-num}

\bibliography{cas-refs}



\end{document}